\documentclass{jfm}

\usepackage{graphicx}
\usepackage{newtxmath}
\usepackage{amsmath}
\usepackage{natbib}
\usepackage{hyperref}
\hypersetup{
    colorlinks = true,
    urlcolor   = blue,
    citecolor  = blue,
}

\newcommand{\RomanNumeralCaps}[1]
\linenumbers

\author{Rahul Agrawal\aff{1}
\corresp{\email{rahul29@stanford.edu}},
  Sanjeeb T. Bose\aff{2},\aff{3} 
  Kevin P. Griffin\aff{4} \and
  Parviz Moin\aff{1}}

  \affiliation{\aff{1}
 Center for Turbulence Research, Stanford University, CA
\aff{2} Cadence Design Systems, San Jose, CA
\aff{3} Institute for Computational and Mathematical Engineering, Stanford University, CA
\aff{4}National Renewable Energy Laboratory, Golden, CO}

\title{An extension of Thwaites' method for turbulent boundary layers}

\begin{document}
\maketitle

\begin{abstract}

\noindent
Thwaites (1949) developed an approximate method for determining the evolution of laminar boundary layers.  The approximation follows from an assumption that the growth of a laminar boundary layer in the presence of pressure gradients could be parameterized solely as a function of a flow parameter, $m = \theta^2/\nu \frac{dU_e}{ds}$, thus reducing the von Karman momentum integral to a first-order ordinary differential equation. This method is useful for the analysis of laminar flows, and in computational potential flow solvers to account for the viscous effects. However, for turbulent flows, a similar approximation for turbulent boundary layers subjected to pressure gradients does not yet exist. In this work, an approximate method for determining the momentum thickness of a two-dimensional, turbulent boundary layer is proposed. It is shown that the method provides good estimates of the momentum thickness, when compared to available high-fidelity simulation data, for multiple boundary layers including both favorable and adverse pressure gradient effects, up to the point of separation. In the limit of high Reynolds numbers, it is possible to derive a criterion for the onset of separation from the proposed model which is shown to be in agreement with prior
empirical observations (Alber, \textit{$9^{th}$ Aerospace Sciences Meeting, 1971}).  The sensitivity of the separation location with respect to upstream perturbations is also analyzed through this model for the NASA/Boeing speed bump and the transonic Bachalo-Johnson bump. 

%

\end{abstract}

\begin{keywords}

\end{keywords}

\section{Introduction} 
Often in the early stages of the engineering design processes, viscous effects are 
estimated using boundary layer integral methods.  Common elements in many such engineering flows are that boundary layers are turbulent and subjected to strong pressure gradients - both 
favorable and adverse. A pressure gradient effect can alter the transition location, or even cause the flow to separate, leading to significant changes in the flow patterns. \citet{clauser1954turbulent,clauser1956turbulent} studied turbulent boundary layers and defined a parameter, $\beta$ that quantifies the relative strength of the pressure gradient in relation to the skin friction across the boundary layer, 
\begin{equation}
    \beta = \frac{\delta^*}{\rho u_{\tau}^2} \frac{dP}{ds}
\end{equation}
where $\delta^{*}$ is the displacement thickness, $dP/ds$ is the pressure gradient and $u_{\tau}$ is the friction velocity. This parameter has since been used in the analyses of equilibrium and non-equilibrium boundary layers. 
In fact, in the presence of strong, and/or prolonged favorable pressure gradients, the flow may relaminarize \citep{sreenivasan1982laminarescent}; thus regaining an ordered structure with a reduction in Reynolds stresses. On the other hand, in the presence of adverse pressure gradients, it is possible for the flow to eventually separate \citep{simpson1981review,simpson1983model}, often associated with periods of rapid growth of the boundary layer. An adverse pressure gradient significantly energizes the outer flow \citep{knopp2021experimental}, and increases the momentum deficit compared to the freestream value. In fact, the effect of a pressure gradient is not only local but strongly dependent on its history \citep{bobke2017history,parthasarathy2023family,prakash2023turbulent}. \citet{wei2017integral} derived the relationships between the turbulent shape factor and the quantities at the edge of the boundary layer of an adverse pressure gradient boundary layer and demonstrated that the boundary layer thicknesses, its growth rate, and the resulting skin friction are affected by upstream effects. Other experimental and computational investigations include the works of  \citet{harun2013pressure,skote1998direct,knopp2015investigation,kitsios2017direct,romero2022properties}; these studies have indicated that the adverse pressure gradient energizes both the large-scale and the smaller-scale structures, while also modulating the impact of the large scales on the smaller scales. 

In the limit of thin boundary layers, the growth rate of the momentum thickness is related to the pressure gradient as follows (via the Karman momentum integral, see \citet{white2008fluid} for more details), 
\begin{equation}
    \frac{d \theta}{d s} = \frac{C_f}{2} + (2 + H) \frac{\theta}{U^2_e} \frac{d P_e}{d s} 
    \label{eqn:vki}
\end{equation}
where $\theta$ is the momentum thickness, and $C_f$ is the skin-friction along the streamwise coordinate. $H$ is the shape factor which is the ratio of the displacement and the momentum thickness lengths of the boundary layer. However, this equation, in itself, is not particularly useful as a predictive tool and requires the knowledge of the skin friction and shape function distributions, which may be unknown \emph{a priori}. 

For laminar boundary layers, \citet{thwaites1949approximate}, and later \citet{curle1957approximate},  and \citet{dey1990extension} fitted data for various flows, including those with pressure gradients and flows undergoing separation, to derive an approximate expression for the growth of the momentum thickness for a given inviscid (freestream) velocity distribution. This analysis coupled with the  \citet{falkner1931lxxxv} analysis in the viscous layer provides a complete description of the mean velocity profile of the corresponding laminar flow. However, a similar analysis for turbulent boundary layers is more complex and lacking. \citet{weber1978boundary} developed a semi-empirical method for calculating the thickness of turbulent boundary layers with arbitrary pressure gradients for the AFOSR-IFP-Stanford conference \citep{coles1968computational,kline1969computation}. However, his approach relied on implied correlations between the skin friction, shape factor, and an inviscid pressure gradient parameter, some of which may be unknown, especially in complex engineering flows.  \citet{das1987numerical}, \citet{das1986integral}, \citet{kalkhoran1986integral} developed integral approaches for measuring skin friction in turbulent boundary layers. A drawback of \citet{das1987numerical}'s and \citet{das1986integral}'s approaches included the use of multiple correlations while also not accounting for history effects (through appropriate streamwise derivatives). \citet{kalkhoran1986integral} made equilibrium assumptions for the wake and the pressure gradient even for non-equilibrium boundary layers. A comprehensive review of these approaches can be found in \citet{das2004comparison}. 


From a practical standpoint, the effects associated with pressure gradients pose significant modeling challenges for near-wall models that are employed to predict wall-shear stress in scale-resolving methods. In the past, various modeling efforts have been pursued \citep{balaras1996two,wang2002dynamic,duprat2011wall,hickel2013parametrized,park2014improved} to include the non-equilibrium effects due to pressure gradients. However, these models have not shown significant differences in their predictive performance in comparison to the equilibrium wall model \citep{cabot2000approximate,kawai2012wall} across a wide range of flows. \cite{park2014improved} developed a non-equilibrium wall model while including the full two-dimensional RANS equations, which adds to the computational complexity. This model was shown to have limited improvements over the equilibrium wall model in the flow past a hump \citep{park2017wall}. \citet{duprat2011wall} did not account for convective terms in their non-equilibrium wall model while adjusting the RANS eddy viscosity to account for pressure gradients. \cite{hickel2013parametrized} accounts for both convective, and pressure gradient terms, but inconsistently uses an equilibrium flow-based eddy-viscosity model.

It is apparent from the aforementioned efforts that further developments in the understanding of the growth of turbulent boundary layers under the action of a pressure gradient are needed for improving potential flow solvers and for near-wall models to account for pressure gradient effects. The objective of this paper is to investigate whether a simple extension to the original Thwaites method for turbulent flows can be derived based on the assimilation of a limited set of high-fidelity simulation data.  The rest of this article is organized as follows, Section II describes the original Thwaites method for laminar flows, and Section III provides the proposed extension to turbulent flows. Section IV describes the simulation database used in this study. Section V describes the model coefficients and their validation in several different flows. Section VI discusses an application of the model in terms of the sensitivity of the separation parameter to flow perturbations. 
Concluding remarks are offered in Section VII.

\section{\label{sec:thwaites} Thwaites method for laminar boundary layers}

In this section, a brief summary of the method of \citet{thwaites1949approximate} for approximating the momentum thickness of laminar boundary layers is provided. Recall, that for thin boundary layers in an incompressible flow, the Von Karman integral equation (\citep{white2008fluid}) is written as, 
\begin{equation}
    \frac{d \theta}{d s} + (2 + H) \frac{\theta}{U_e} \frac{d U_e}{d s} = \frac{C_f}{2}
\end{equation}
where in the $n-s$ coordinate system, $s$ is the streamwise coordinate, and $n$ is the wall-normal coordinate. $\delta^{*}$ and $\theta$ are the displacement and momentum thickness lengths of the boundary layer, respectively, and $H = \delta^*/\theta$ is the shape factor. The following relations also hold under the thin boundary layer approximation,
\begin{equation}
  \nu  \frac{\partial^2 U}{\partial n^2} \bigg|_{n=0} = \frac{\partial P_e}{\partial s} = - U_e \frac{\partial U_e}{\partial s}
\end{equation}
where $P_e$ and $U_e$ are the values of the pressure, and the streamwise velocity at the edge of the boundary layer, and $n=0$ refers to the wall. Using these relations, Thwaites derived that 
\begin{equation}
   2 Re_{\theta} \frac{d \theta}{d s} =  2\left(\left(2 + H\right) m + \chi\right)  = L
   \label{eqn:thwaitesoriginal}
\end{equation}
where $m$ is the Holstein-Bohlen pressure gradient parameter, $m = \frac{\theta^2}{U_e} \frac{\partial^2 U }{\partial n^2} \big|_{n=0} $,
and $\chi = \frac{\theta}{U_e} \frac{\partial U}{\partial n} \big|_{n=0}$. 
It should be noted that in deriving Equation \ref{eqn:thwaitesoriginal}, the assumption of a laminar flow has not yet been invoked. For laminar boundary layers, however, Thwaites further postulated that $L = L(m)$, or that the effect of the pressure gradient primarily determines the growth rate of a laminar boundary layer. Specifically, the linear fit proposed in Thwaites' original work is 
\begin{equation}
    L \approx L(m)^{Thwaites} \approx 0.45 + 6m
    \label{eqn:thwaitesmodel}
\end{equation}

Rearranging the model fit, an approximate closed-form expression of the momentum thickness can be written as, 
\begin{equation}
    \theta^2(s) \approx 0.45 \frac{\nu}{U^6_e} \int^s_0 U^5_e(r) dr 
\end{equation}
Further, Thwaites fitted experimental data to suggest that a laminar flow may separate when $m \approx 0.09$. Other similar empirical thresholds have been proposed by \citet{stratford1959prediction}, and by \citet{curle1957approximate}. More details can be found in \citet{horton1968laminar}. 
\citet{curle1957approximate}, and later \citet{dey1990extension} suggested slightly different model coefficients that resulted in a better agreement with the Blasius and Falkner-Skan type solutions. These methods have been used in potential flow solvers to account for viscous effects in laminar flows. 

\section{\label{sec:fit} Extension to turbulent flows}
Unlike in laminar flows, the inner and outer layers of a turbulent boundary layer respond differently to external pressure gradients. Further, a thin internal layer develops near the wall in turbulent flows due to the pressure gradient \citep{uzun2021high,parthasarathy2023family}, and the amplitude of the small scales near the wall gets modulated due to the larger ones \citep{knopp2015investigation,romero2022properties}. A general method for predicting the momentum thickness, $\theta$, similar to that described in the previous section is hence lacking for general, non-equilibrium turbulent flows. 

Consider a zero pressure gradient turbulent boundary layer at a high Reynolds number. The growth rate of the momentum thickness of a turbulent boundary layer is higher than that of an equivalent Blasius laminar boundary layer ($\theta/s \sim Re_s^{-1/2}$). Reconciling this fact with the fact that Equation \ref{eqn:thwaitesoriginal} still holds for turbulent flows, the approximation made by Thwaites in Equation \ref{eqn:thwaitesmodel},  $L = L(m)$ must not hold for turbulent boundary layers. Since a boundary layer at a higher Reynolds number is less sensitive to the effects of pressure gradients \citep{vinuesa2018turbulent}, the growth rate can not depend only on the pressure gradient (as it did in the laminar case). Therefore, it is proposed that the growth rate of the momentum thickness may also depend directly on the Reynolds number. Equation \ref{eqn:thwaitesoriginal} can be re-expressed as, 
\begin{equation}
    2\frac{d \theta}{d s} =  2\frac{(2 + H) m}{ Re_{\theta} } + \frac{2\chi}{Re_{\theta}}. 
    \label{eqn:asymptotic0}
\end{equation}
A Taylor's series expansion of $2 {d\theta}/{ds}$ in terms of $m/Re_{\theta}$ and $1/Re_{\theta}$ gives, 
\begin{equation}
    2\frac{d \theta}{d s} = C_{Re, \infty} + C_m \frac{m}{Re_{\theta}} + C_c\frac{1}{Re_{\theta}} + C_{m,2} \left(\frac{m}{Re_{\theta}}\right)^2 + C_{c,2}\left(\frac{1}{Re_{\theta}}\right)^2 +C_{m,c}\left(\frac{1}{Re_{\theta}}\right) \left(\frac{m}{Re_{\theta}}\right)+ \cdots,
    \label{eqn:asymptotic1}
\end{equation}

\noindent
where $C_{(\cdot)}$ are the constant coefficients of the Taylor expansion about some intermediate values of $m$ and $Re_\theta$. In the limit of a large Reynolds number, $Re_{\theta}$, such that $m/Re_{\theta}, \; 1/Re_{\theta} \ll 1 $, the linear truncation is given by
\begin{equation}
    2\frac{d \theta}{d s} \approx C_{Re, \infty} + C_m \frac{m}{Re_{\theta}} + C_c\frac{1}{Re_{\theta}}.
    \label{eqn:trunc_expansion}
\end{equation}

\noindent
This model can be thought to be based on the assumptions that the growth of the boundary layer and that the effect of the pressure gradient decreases with the Reynolds number. An interpretation of the linear model is that the effect of the pressure gradient, and the Reynolds number are only weakly coupled. Rewriting Equation \ref{eqn:trunc_expansion} in terms of the definition of $L$ in Equation \ref{eqn:thwaitesoriginal}, the model form for $L$ can be written as, 
\begin{equation}
    L = L(m, Re) \approx C_c + C_m m + C_{Re, \infty} Re_{\theta}
\end{equation}

\noindent
With this approximation, the growth rate of the momentum thickness is
\begin{equation}
\begin{split}
    \frac{d}{d s} (U^{C_m}_e \theta^2) \approx  \nu C_c U^{C_m-1}_e   +  C_{Re, \infty} {U^{C_m}_e \theta }.
\end{split}
\label{eqn:mymodel}
\end{equation}
Unlike laminar boundary layers, this system does not have a closed-form solution for arbitrary pressure gradients. However, the equation is an ordinary 
differential equation that can be integrated numerically along the streamwise direction if the inflow and the freestream conditions are known.

\subsection{Zero pressure gradient boundary layers}
The special case of a zero pressure gradient (ZPG) boundary layer is now considered ($U_e =$ constant, $m=0$). Under this limit
\begin{equation}
\begin{split}
     \frac{d \theta }{d s} \approx  \frac{C_c}{2} \frac{1 }{ Re_\theta}   +  \frac{C_{Re, \infty}}{2}
\end{split}
\end{equation}
The analytical solutions to this system suggest that the growth rate is slightly sublinear (and higher than $s^{0.5}$) which is in line with experimental evidence. From this model equation, it may be hypothesized that the turbulent contribution (contribution from the $C_{Re, \infty}$ term) becomes relatively dominant when $Re_{\theta} > C_c/C_{Re, \infty} \approx 600 $ (the values of $C_c$ and $C_{Re, \infty}$ are provided later in Section 5), which implies $ Re^{zpg}_{\tau} \geq  200$.

\subsection{Constant pressure gradient boundary layers}
In the case of a constant (or a slowly varying) externally applied constant pressure gradient, the Bernoulli equation evaluated in the inviscid freestream gives 
\begin{equation}
    \frac{d P}{d s} = - U_e \frac{d U_e}{d s} = - \Lambda
\label{eqn:dpdxdudx}
\end{equation}
Then at a far downstream distance from the inflow, the freestream velocity can be approximated as $U_e (s) \sim \sqrt{ U^2_0  - 2 \Lambda s} $ where $U_0$ is the freestream velocity at $s=0$. Invoking Equation \ref{eqn:mymodel} gives 
\begin{equation}
    \frac{d \theta}{d s} \approx \frac{C_c}{2} \frac{\nu}{U_e \theta} + \frac{C_{Re, \infty}}{2} + \frac{C_m}{2} \frac{\theta }{U^2_e} \frac{d P_e}{d s}.
    \label{eqn:mymodel2}
\end{equation}

The third term scales as $ \theta / s $, thus the effect of a constant pressure gradient is to affect the growth of the boundary layer linearly.

. 


\section{\label{sec:Data} Database of pressure gradient flows}

It has been well established that for turbulent boundary layers, the presence of strong pressure gradients leads to ``history effects", or that two boundary layers experiencing different upstream pressure gradients may respond differently to the action of the same pressure gradient even when the Reynolds number ($Re_{\theta}$ or $Re_{\tau}$) is held constant \citep{bobke2017history,devenport2022equilibrium,vinuesa2017revisiting,nickels2004inner}. This is because under the influence of the pressure gradient, the near-wall, and outer layer scales of motion respond at different spatial, and temporal scales \citep{knopp2021experimental}. To account for many of these effects, a comprehensive database of turbulent boundary layers is used in this work. All of the data used is from numerical simulations for ease of accessing continuous data, for computing streamwise derivatives and integrals. Relevant details of the datasets are given below. Table \ref{table:data-table} summarizes the variation in the Reynolds number and the Clauser parameter for the boundary layers considered. It is noteworthy that \citet{goctransoniccrm} reported similar skin-friction Reynolds numbers across the wing of a transonic aircraft model as those observed in this database. \\

\begin{table}
\begin{tabular}{ p{8cm}p{2cm}p{2cm}}
Dataset & Range of $Re_{\tau}$ & Range of $\beta$\\
\hline
ZPG (\citet{eitel2014simulation})    & [500,2700] & 0\\
APG (\citet{bobke2017history})  & [300,1000] & [1, 5]\\
Airfoils (\citet{vinuesa2018turbulent},  \citet{tanarro2020effect})  & [100,700] & [1, 15] \\
FPG/APG SBSE (\citet{uzun2021high}) & [400,1200] & [-2, 15] \\
FPG/APG BJ (\citet{uzun2019wall})  & [1500,4000] & [-6, 6] \\
\end{tabular}
\caption{The variation in friction-based Reynolds number and Clauser parameters in the datasets considered in this work. For the flows that experience separation (FPG/APG SBSE and FPG/APG BJ), only pre-separation data is considered. Since both favorable and adverse pressure gradient boundary layers are included, the Clauser parameter ranges from a negative to a positive value.  }
\label{table:data-table}
\end{table}

\begin{itemize}
    \item \textbf{ZPG Eitel}: \citet{eitel2014simulation} performed wall-resolved large eddy simulations (LES) of a spatially developing zero pressure-gradient boundary layer up to $Re_{\theta} = 8300$. 
    
    \item \textbf{APG Bobke}: \citet{bobke2017history}  performed wall-resolved LES of five different boundary layers at varying Reynolds numbers and strength of pressure gradients. Two of these five boundary layers were maintained at nearly constant Clauser parameters, $\beta =1, \; \mathrm{and} \; 2$ while the other three boundary layers followed a power law for the streamwise velocity at the edge of the boundary layer, $U_e \sim (x-x_0)^m$ with $m=-0.13, -0.16, \; \mathrm{and} \; -0.18$. 
    
    \item \textbf{Airfoils}: \citet{vinuesa2018turbulent} and \citet{tanarro2020effect} simulated the flow over two different NACA airfoils, namely NACA 4412 and 0012 respectively at varying chord-length-based Reynolds number and angles of attack. In this work, we consider three particular cases, NACA 0012 airfoil at $Re_c = 0.4 \mathrm{Mil}$, $\alpha=0.0^o$ deg., and NACA 4412 airfoil at $Re_c = 0.1 \mathrm{Mil}, \; 1 \mathrm{Mil}$, $\alpha=5.0^o$ deg.
    
    \item \textbf{FPG/APG SBSE}: The problem of predicting smooth body separation using both Reynolds averaged methods (RANS) and scale-resolving methods in CFD has remained a challenge. \citet{uzun2021high} performed a quasi-DNS of the flow over a smooth Gaussian bump as it experienced both a favorable pressure gradient, followed by an adverse pressure gradient, and thereafter a turbulent separation. These results were in agreement with the experiments of  \citet{williams2020experimental}. Later, \citet{agrawal2022non,whitmore2022brief}  performed wall-modeled large-eddy simulations of this flow and accurately predicted the flow separation.
    
    \item \textbf{FPG/APG BJ}: In aerospace applications, such as the flow over the transonic Common Research Model (CRM) aircraft \citep{rivers2010experimental}, weak compressibility effects affect the flow pattern due to the shock-formation, and the consequent flow separation at high Mach numbers. \citet{uzun2019wall} performed high-fidelity simulations of the experiments of \citet{bachalo1986transonic} that studied the flow over a hump on a cylinder experiencing shock and adverse pressure-gradient induced flow separation. 
\end{itemize}

\vspace{0.2in}
\noindent
The undetermined model coefficients, $C_c, \; C_m, \; \mathrm{and} \; C_{Re, \infty}$, are ascertained from fitting the simulation data from the zero pressure gradient 
and adverse pressure gradients of \cite{eitel2014simulation} and \cite{bobke2017history}, respectively.  The remainder of the simulation database is used to 
assess the generality of the proposed model and fits.

\section{\label{sec:implication} Model Coefficients and Validation }

\subsection{Fitting model coefficients }

\begin{figure}
    \centering
    \includegraphics[width=1\textwidth]{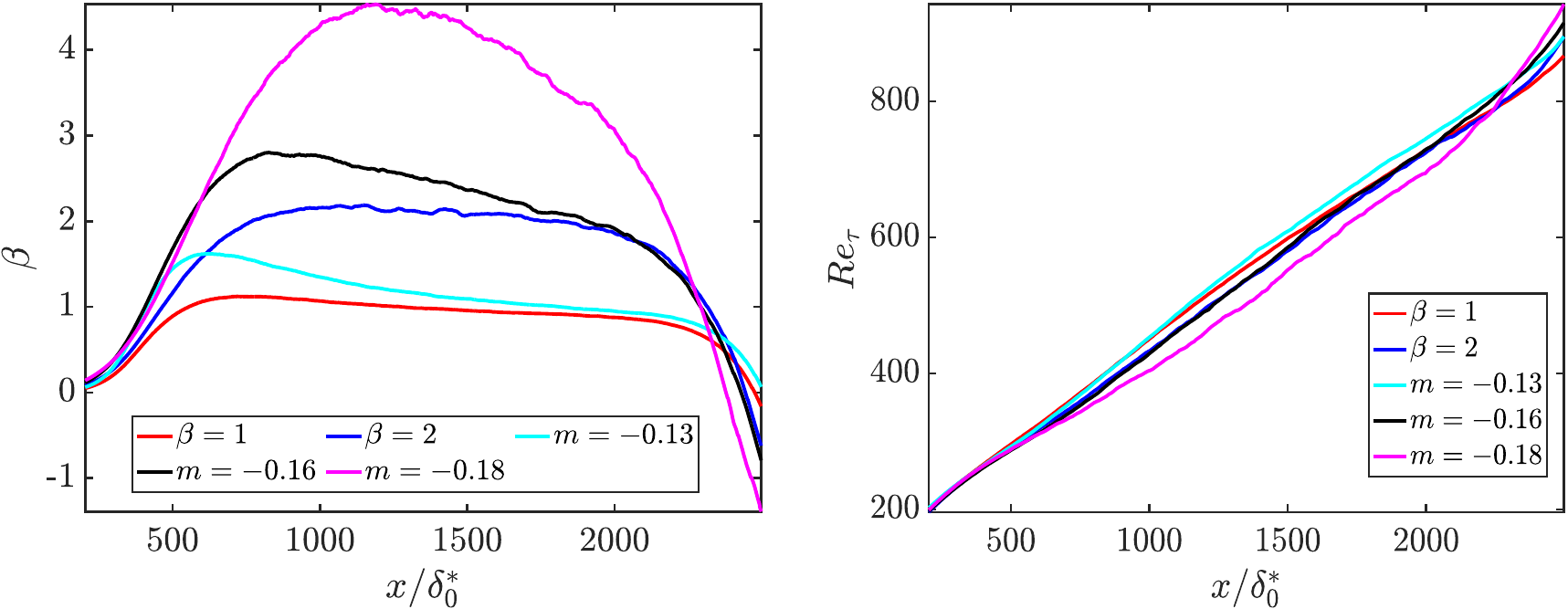} 
    \caption{The distribution of the (a) Clauser parameter, $\beta$, and the (b) friction Reynolds number, $Re_{\tau}$, for the boundary layers considered in  \citet{bobke2017history}}.
    \label{fig:bobkebeta}
\end{figure}

\begin{figure}
    \centering
    \includegraphics[width=0.5\textwidth]{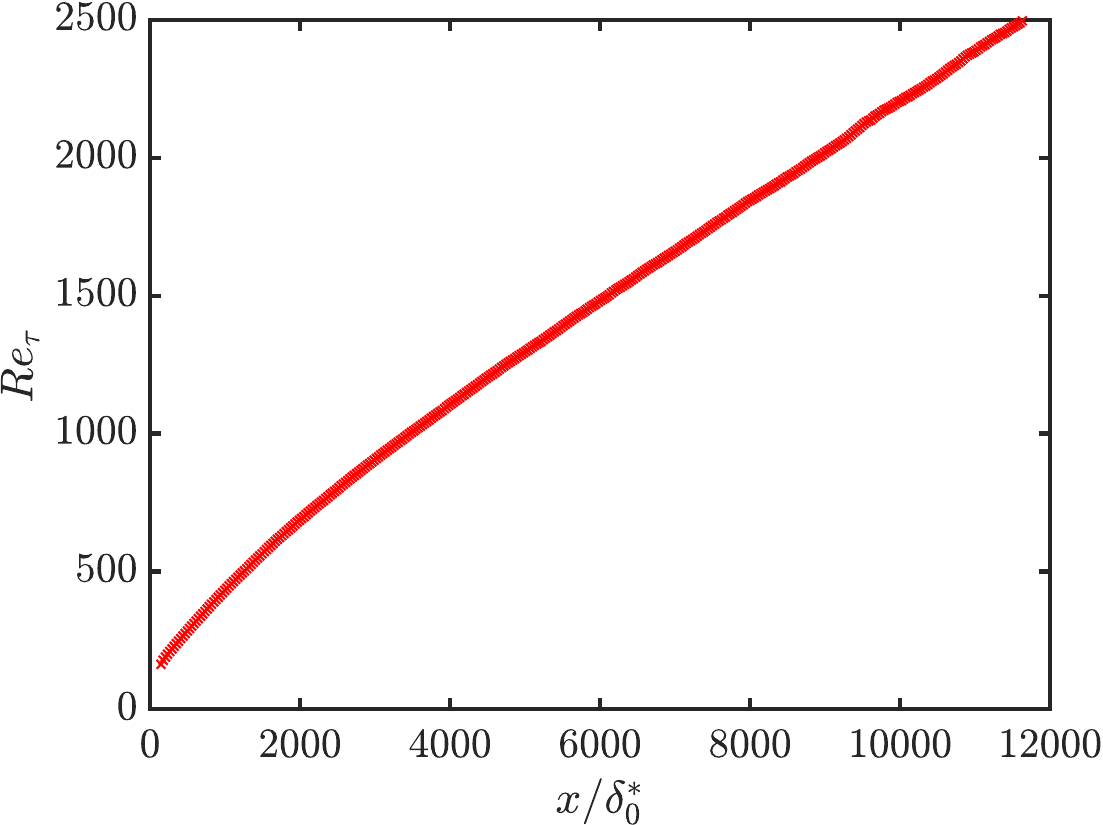}
    \caption{The distribution of the skin-friction based Reynolds number, $Re_{\tau}$, for the zero-pressure gradient boundary layer studied in  \citet{eitel2014simulation}.}
    \label{fig:zpgRe}
\end{figure}

The model coefficients $C_c, \; C_m, \; \mathrm{and} \; C_{Re, \infty}$ are determined in two stages. Since $m = \frac{\theta^2}{U_e} \frac{\partial^2 U }{\partial n^2} \big|_{n=0} = 0$ for a zero pressure-gradient boundary layer, the coefficients $C_c$ and $C_{Re, \infty}$ are determined from the data of \citet{eitel2014simulation} alone using a least-squares fit (with bisquare weights).  Then $C_m$ is fitted using the wall-resolved LES data of boundary layers under slightly dissimilar pressure gradients \citep{bobke2017history}. The numerical values of the coefficients obtained in this way are, $C_c =1.45 $, $C_{Re, \infty} = 0.0024 $ and $C_m = 7.20 $ with the corresponding $95\%$ confidence intervals $C_c \in [1.43,1.47]$, $C_{Re, \infty} \in[0.0023,0.0025]$ and $C_m \in[7.18,7.21]$ respectively.  Once the three coefficients are fitted, they are kept constant as the fit is tested on other flows.  Figures \ref{fig:bobkebeta} and \ref{fig:zpgRe}  show the variation in the Reynolds number and the Clauser parameter for the zero-pressure gradient boundary layer, and the five adverse pressure gradient boundary layers. The adverse pressure gradient flows have slightly different pressure-gradient distributions which lead to different growth rates of the boundary layers as the outer scales are energized differently. \\

\begin{figure}
    \centering
    \includegraphics[width=1.0\textwidth]{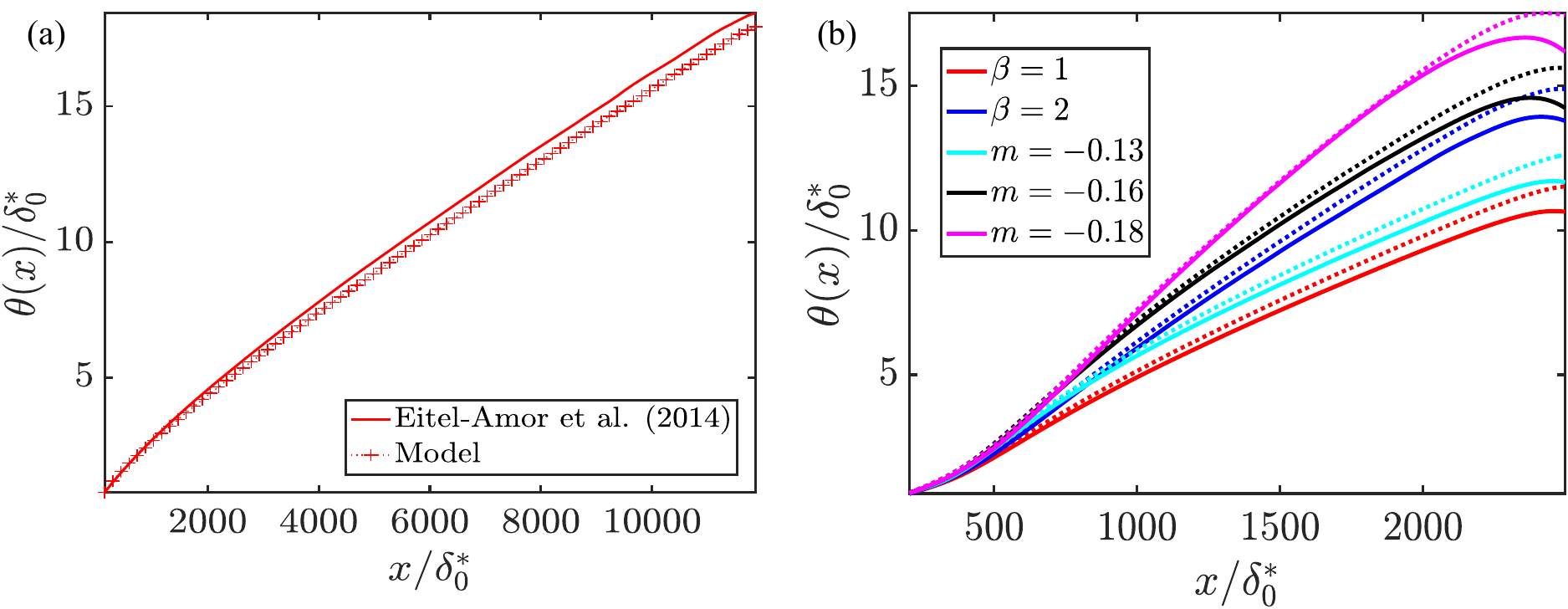}
    
    \caption{Fit of $\theta$ vs the streamwise distance $x/\delta^*_0$. Subfigure (a) is a model fit for the zero pressure gradient boundary layer in \citet{eitel2014simulation} and subfigure (b) is fit for the adverse pressure gradient boundary layers in \citet{bobke2017history}. In subfigure (b) solid lines denote the values obtained from simulations of \citet{bobke2017history} and dotted lines refer to the model fit.   }
    \label{fig:zpgapg}
\end{figure}

\noindent
Using Equation \ref{eqn:mymodel}, the variation in $\theta$ vs. $s$ is compared against the values obtained from mean velocity profiles using the method of \citet{griffin2021general}. Figure \ref{fig:zpgapg} shows the model fit for the zero-pressure gradient boundary layer, and the adverse pressure gradient boundary layers, clearly the growth rate of $\theta$ is reasonably predicted with minor discrepancies near the outlet of the domain.\\

\begin{figure}
    \centering
    \includegraphics[width=0.5\textwidth]{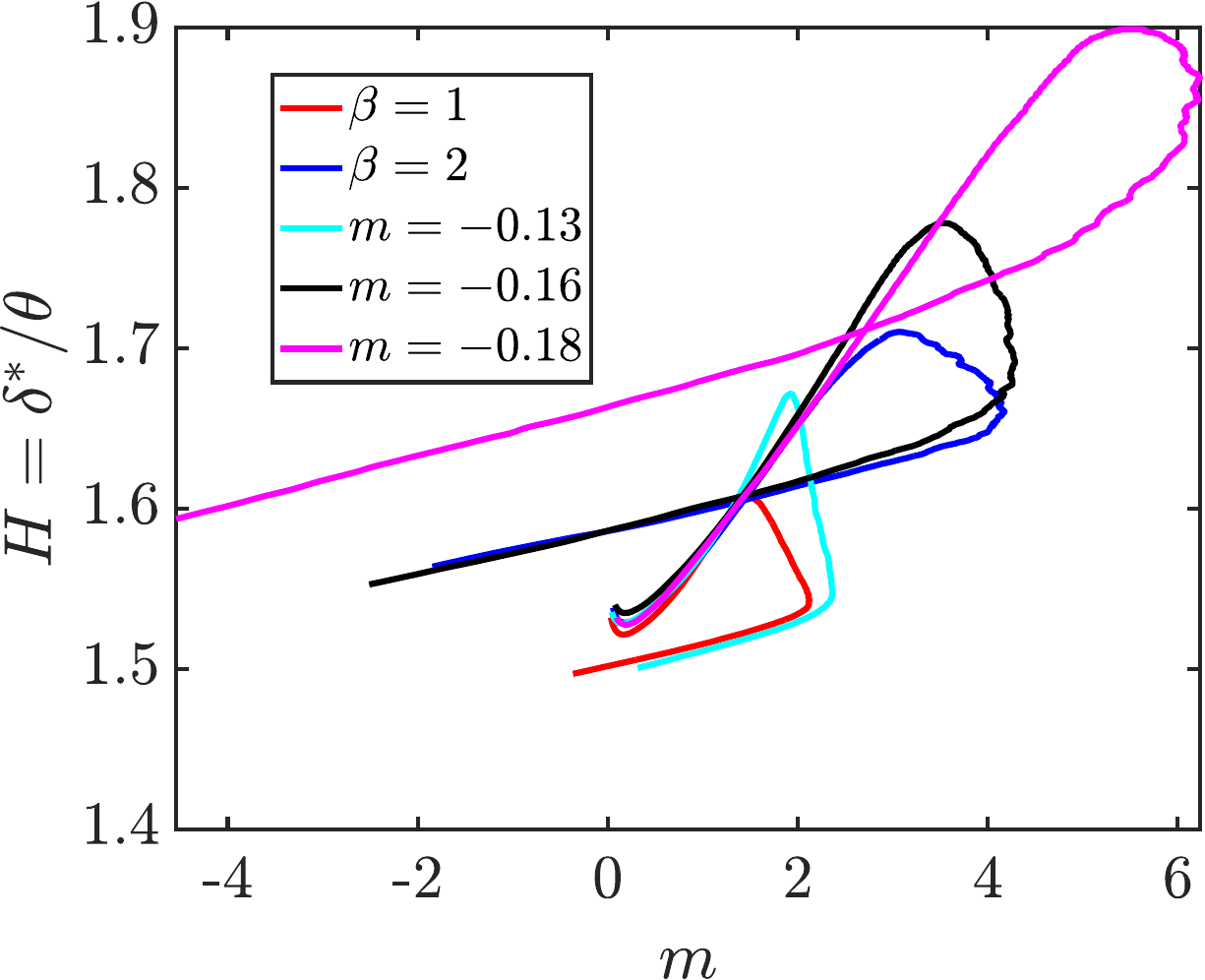} 
    \caption{The variation of the shape factor $H$ and Holstein-Bohlen parameter $m$ for the five adverse pressure gradient boundary layers in \citet{bobke2017history}.  }
    \label{fig:bobkeHvsm}
\end{figure}

\noindent
It is worth noting that unlike laminar boundary layers, turbulent boundary layers do not exhibit a universal relation between the shape factor, $H$, and the Holstein-Bohlen pressure gradient parameter, $m$. Figure \ref{fig:bobkeHvsm}
shows that for the boundary layers in \citet{bobke2017history}, the shape factor varies significantly across the cases, and hence a universal fit of $H$ with $m$ is not possible. Although not shown, it was also verified that $H$ also does not follow a linear fit in $(m, Re_{\theta})$ space. Laminar potential flow solvers iteratively solve for $\theta$ from Thwaites Method and find $\delta^*$ from the $H = H(m)$ relationship to determine the ``effective body shape" due to flow acceleration. For turbulent flows, since $H = H(m)$ is nonuniversal, this approach is not workable. Appendix \ref{sec:deltastareqn} provides an alternate approach for finding the growth rate of the displacement thickness using boundary layer edge variables.

\subsection{Testing on APG Wing data}

\begin{figure}
    \centering
    \includegraphics[width=0.9\textwidth]{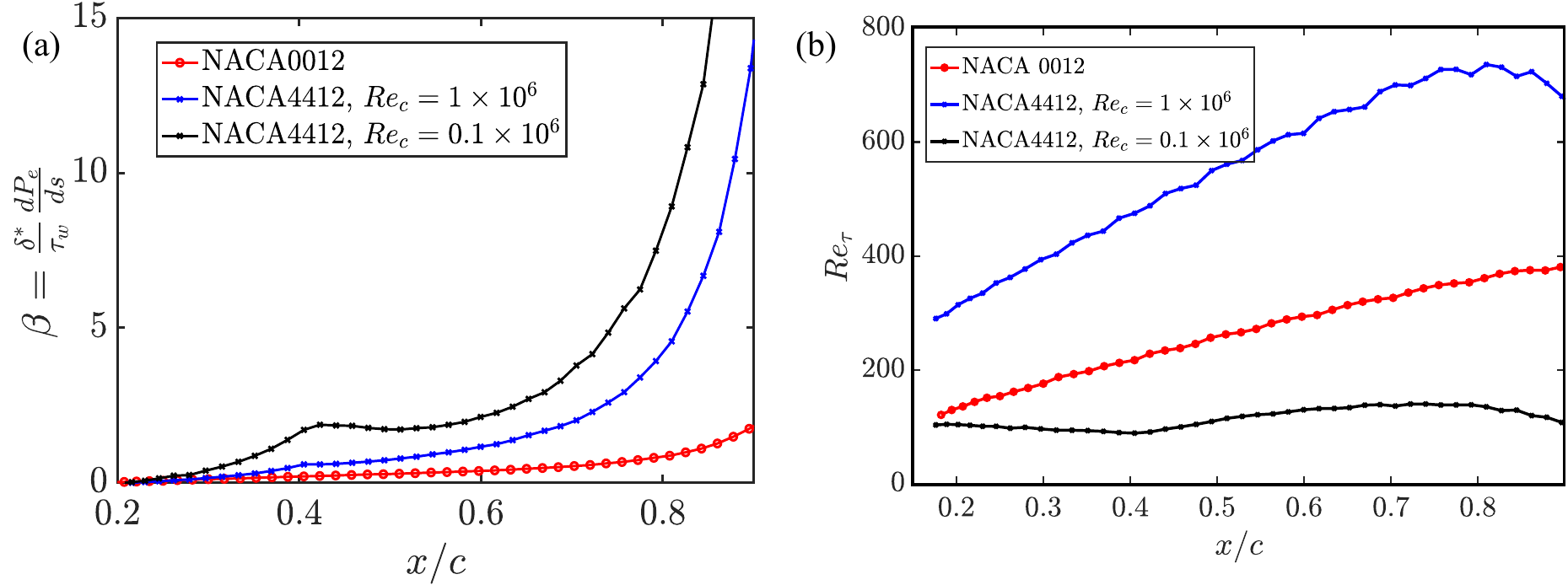}\caption{The distribution of the (a) Clauser parameter $\beta$ and the (b) skin-friction based Reynolds number, $Re_{\tau}$, for a NACA 0012 airfoil at $Re_c = 0.4 \times 10^6$, angle of attack $\alpha = 0^o$ (\citet{tanarro2020effect}) and for a NACA 4412 airfoil at $Re_c = 0.1, 1 \times 10^6$,  angle of attack $\alpha = 5^o$ (\citet{vinuesa2018turbulent}) respectively. }
    \label{fig:wingsretaubeta}
\end{figure}

\begin{figure}
    \centering
    \includegraphics[width=0.5\textwidth]{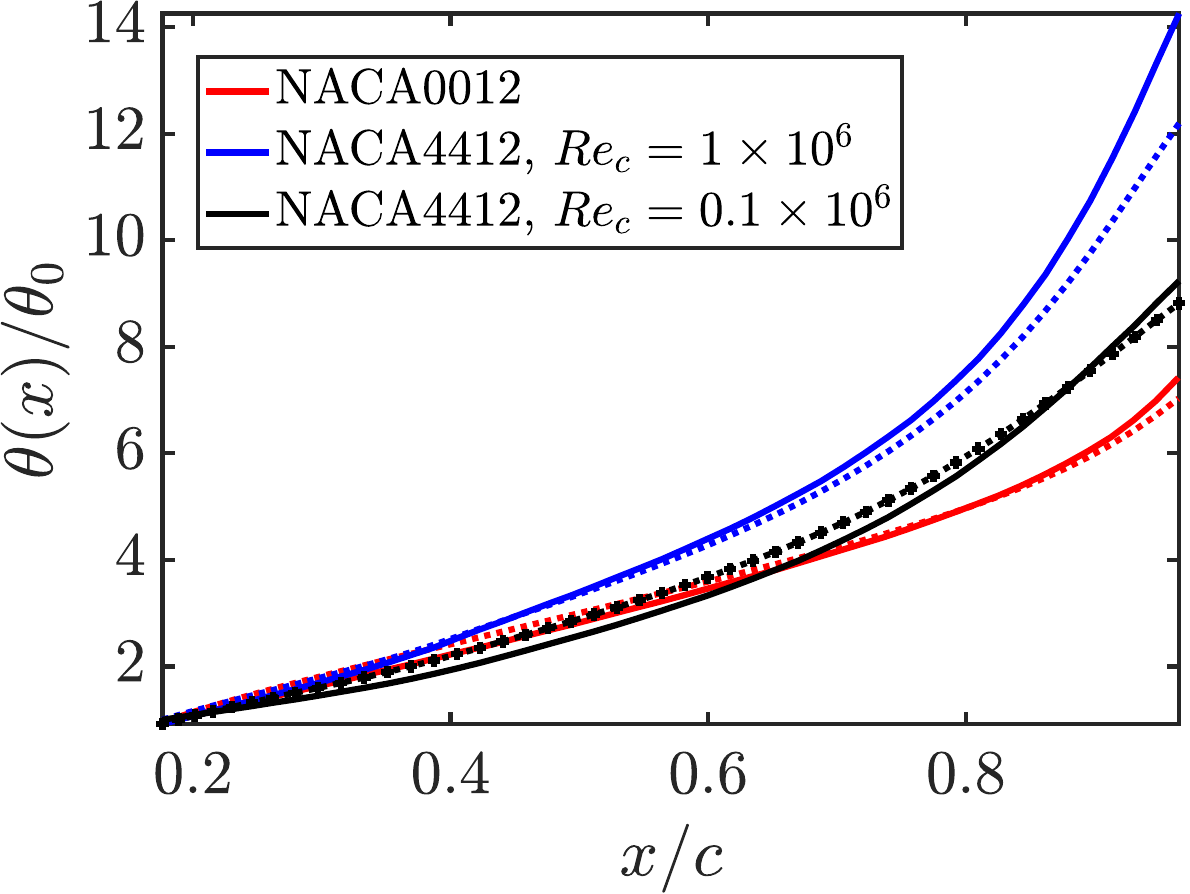} 
    \caption{ A fit of $\theta / \theta_0$ vs the streamwise distance $x/c$ for a NACA 0012 airfoil at $Re_c = 0.4 \times 10^6$, angle of attack $\alpha = 0^o$ (\citet{tanarro2020effect}) and for a NACA 4412 airfoil at $Re_c = 0.1, 1 \times 10^6$,  angle of attack $\alpha = 5^o$ (\citet{vinuesa2018turbulent}) respectively.    }
    \label{fig:wings}
\end{figure}

Now, the model fit is evaluated for two airfoil flows at low angles of attack published in \citet{vinuesa2018turbulent,tanarro2020effect}. Two NACA airfoils at different chord-based Reynolds numbers ($Re_c$) and angles of attack ($\alpha$). A NACA 0012 airfoil at $Re_c = 0.4 \times 10^6, \; \alpha = 0^o$ and NACA 4412 airfoil at $Re_c =0.1 \times 10^6$  and $Re_c =1.0 \times 10^6$, $\alpha = 5^o$ are considered. Note that the dataset available for these airfoils only provides detailed statistics beyond $x/c = 15\%$ as the flow is tripped to become turbulent at $x/c = 0.1$. \\ 

\noindent
For these cases, the pressure gradient distributions are quite different, as quantified by the Clauser parameter in Figure \ref{fig:wingsretaubeta}(a). \citet{goctransoniccrm} recently evaluated that $Re_{\tau}$ over a transonic common research model aircraft (at $\alpha \approx 4^o$) is approximately $Re_{\tau} \approx 800$ at $10 - 20 \%$ chord length (similar to those experienced on the presently considered airfoil flows). Near the trailing edge, the pressure gradient for NACA 4412 airfoils varies quite significantly and is much stronger than effects of the wall-shear stress ($\beta \approx 15$ at the trailing edge) suggesting a strongly non-equilibrium flow. Nonetheless, the fit of the momentum thickness in Figure \ref{fig:wings} is still reasonable in this region, suggesting a simple linear model is capturing the history effects from the pressure gradients. The increased Reynolds number is responsible for the lowered $\beta$ between the two NACA 4412 airfoils; as the lower Reynolds number boundary layer is more strongly affected by the pressure gradient \citep{vinuesa2017revisiting}.

\subsection{Testing on SBSE and BJ data}
\begin{figure}
    \centering
    \includegraphics[width=0.7\textwidth]{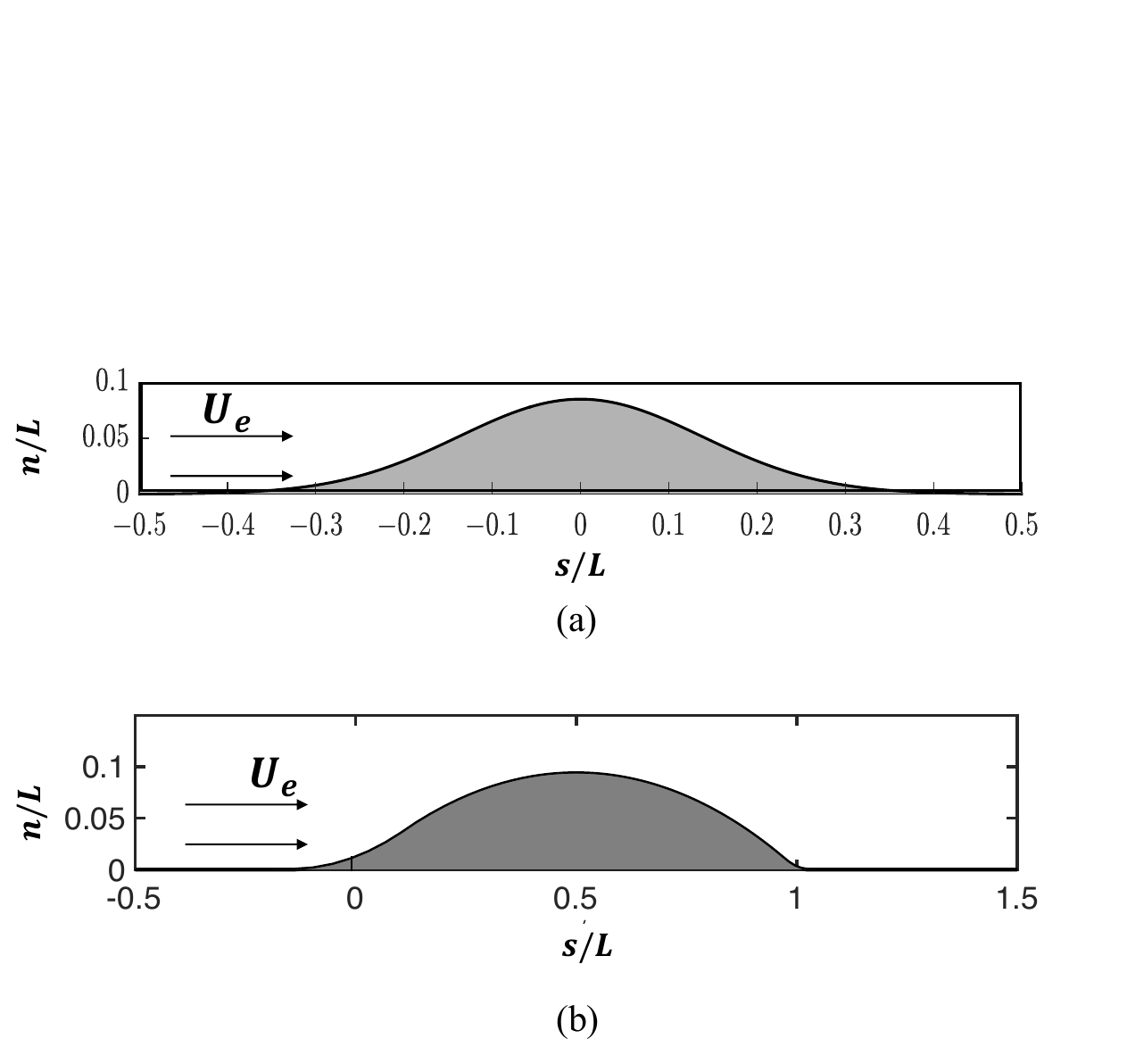}
    \caption{A schematic of the surface geometry over which the turbulent flow separates in the (a) subsonic Boeing speed bump case of \citet{uzun2021high} and (b) transonic flow case of \citet{uzun2019wall}. $L$ and $c$ are the characteristic lengths for the two cases respectively. $U_{e}$ is the freestream flow velocity and the arrows next to it point along the positive streamwise direction. }
    \label{fig:geombumps}
\end{figure}

\begin{figure}
    \centering
    \includegraphics[width=0.9\textwidth]{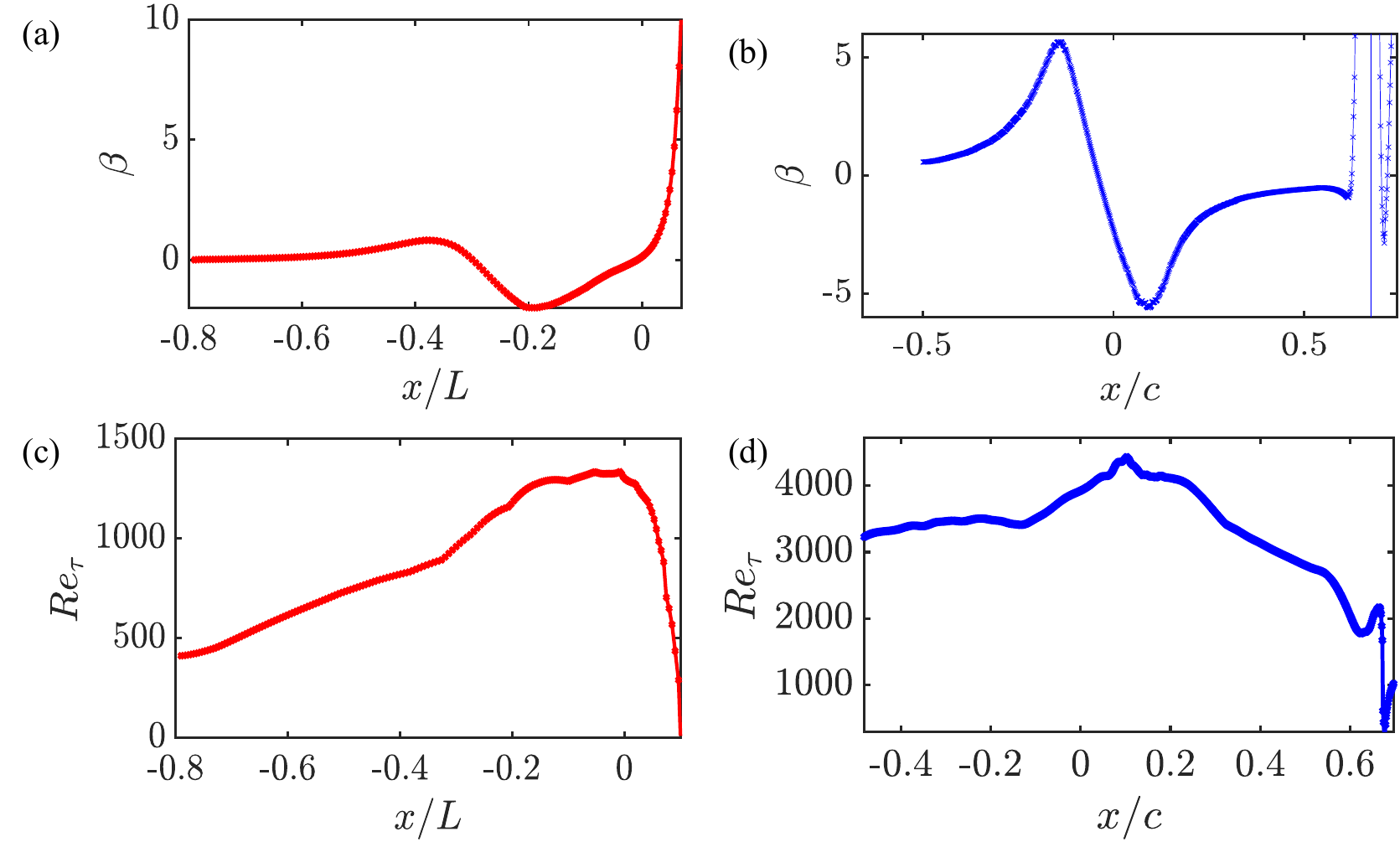}
    \caption{The streamwise distribution of the Clauser parameter $\beta$ (subfigures a, b) and the skin-friction based Reynolds number $Re_{\tau}$ (subfigures c, d)  for the Boeing speed bump (subfigures a, c) in \citet{uzun2021high} and the transonic Bachalo-Johnson bump (subfigures b, d) in \citet{uzun2019wall}. For the transonic flow, the density on the wall is used for defining the Reynolds number. Note that $x/L = 0$ is the position of the apex of the Boeing speed bump, whereas $x/c = 0.5$ is the position of the apex of the transonic bump. }
    \label{fig:sbsebjbetaRetau}
\end{figure}

Smooth body separation refers to the flow condition where neither the separation nor reattachment points of a separation bubble are geometrically imposed. As a result, the flow development history (i.e. the influence of pressure gradients and/or compressibility) affects the tendency of the flow to separate. 
\citet{uzun2021high} recently performed a quasi-DNS of flow over a spanwise-periodic Gaussian bump in which the flow separates on the leeward side of the bump due to the mild, adverse pressure gradient. Similarly, the wall-resolved LES of \citet{uzun2019wall} is an example of shock-induced separation in a transonic flow over a bump. Figure \ref{fig:geombumps} shows a schematic of the bump geometries and the case-setup for both flows. Figure \ref{fig:sbsebjbetaRetau} shows the distribution of the Clauser parameter $\beta$ and the friction Reynolds number $Re_{\tau}$ for these cases. For both flows, an incoming zero-pressure gradient turbulent boundary layer experiences a favorable pressure gradient, which accelerates the flow, before eventually experiencing an adverse pressure gradient that leads to flow separation. The Clauser parameter, Reynolds numbers, and their ``histories" until the point of separation ($x/L \sim 0.1$ for the SBSE, and $x/c \sim 0.65$ for the BJ) are significantly different.

\begin{figure}
    \centering
    {\includegraphics[width=0.9\textwidth]{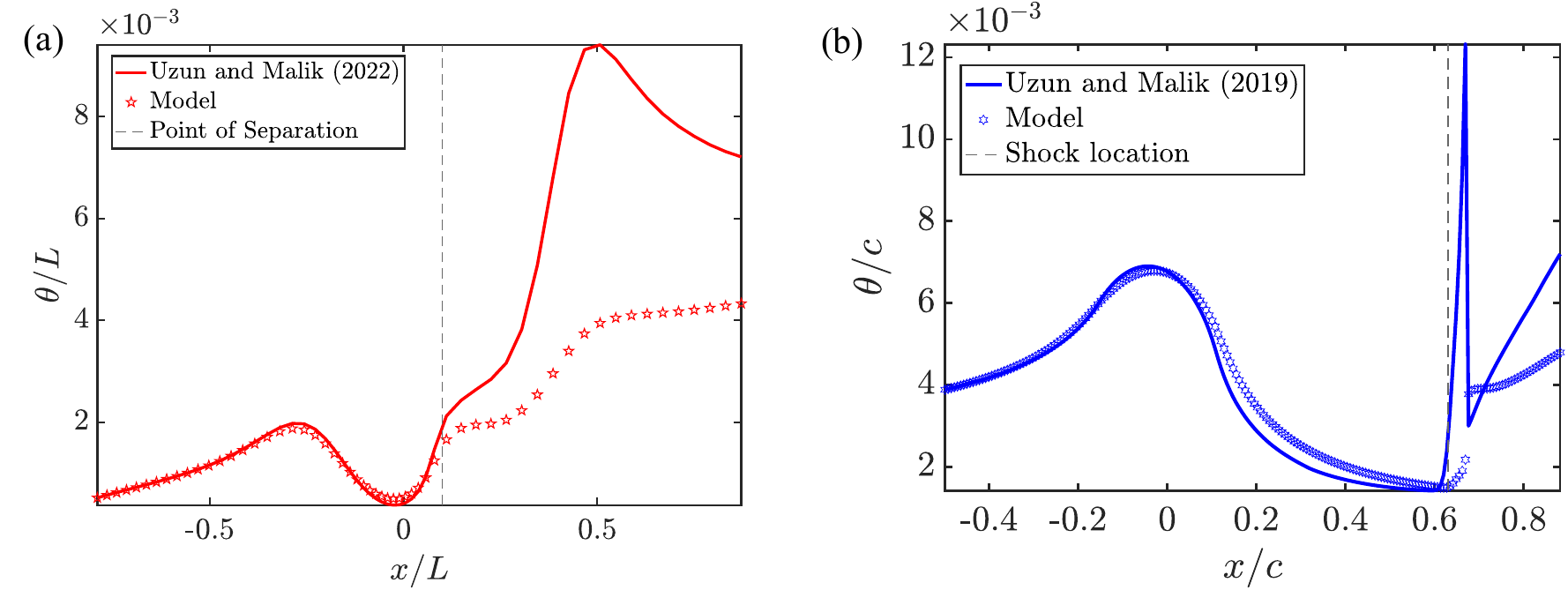}}    
    \caption{Fit of $\theta$ vs the streamwise distance $x$ non-dimensionalized by the respective relevant length scales (inlet displacement thickness of the laminar flow) as defined in the reference simulations. Subfigure (a) is a model fit for the Boeing speed bump in \citet{uzun2021high} and subfigure (b) is a model fit for the transonic Bachalo-Johnson bump in \citet{uzun2019wall}. The solid vertical lines denote the separation point as obtained from the reference data. Note that $x/L = 0$ is the position of the apex of the Boeing speed bump, whereas $x/c = 0.5$ is the position of the apex of the transonic bump. }
    \label{fig:sbsebj}
\end{figure}

As seen in Figure \ref{fig:sbsebj}, the proposed model predicts the momentum thickness for both these flows reasonably up to the point of separation (for the transonic flow, \citet{uzun2019wall} performed their simulations at $Ma = 0.875$ at which the location of the shock, and the separation nearly coincide \citep{bachalo1986transonic}).
The proposed model was derived for incompressible flows, and as such neglects any variations in density in the flow. The transonic flow in \citet{uzun2019wall} exhibited strong density variations; the ratio of the mean density at the edge of the boundary layer and the wall $\approx 2-3$. Finally, it is not surprising for the proposed, linear model to not predict the growth inside the separation bubble as upstream effects of the blockage of the separation bubble are not comprehended by the ODE model employed (see \citet{simpson1981review} for a more comprehensive overview) and the 
thin boundary layer approximation utilized also breaks down at the point of separation.\\

\begin{figure}
    \centering
    {\includegraphics[width=0.9\textwidth]{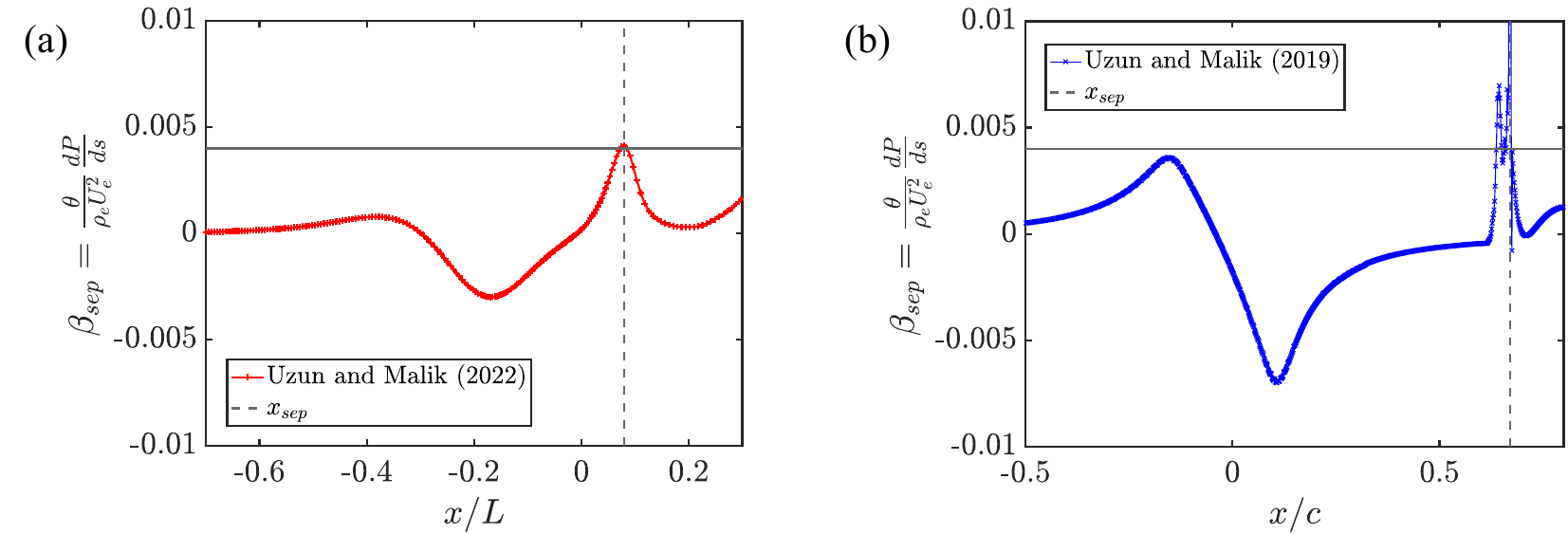}}
    \caption{ The variation in the non-dimensional group $\frac{\theta}{\rho_e U^2_e} \frac{dP}{ds}$ along the streamwise direction for the (a) Boeing speed bump in \citet{uzun2021high} and the (b) transonic Bachalo-Johnson bump in \citet{uzun2019wall}. The dashed vertical lines denote the separation point as obtained from the reference data. Note that $x/L = 0$ is the position of the apex of the Boeing speed bump, whereas $x/c = 0.5$ is the position of the apex of the transonic bump. }
    \label{fig:separationptalber}
\end{figure}

\noindent
However, the proposed model can be used to help predict the conditions under which the separation of the boundary layer is imminent.  To date, there have been numerous attempts to characterize this point of separation empirically.  \citet{alber1971similar} proposed that separation is imminent when $\frac{\theta}{U^2_e} \frac{d P_e}{d s} > 0.004$. The exact value of the threshold was determined empirically for some weakly compressible flows and has been shown to be relatively robust in the presence of weak shocks \citep{alber1973experimental,adair1987characteristics}. As seen in Figure \ref{fig:separationptalber} the value of  $\frac{\theta}{U^2_e} \frac{d P_e}{d s}$ is approximately $0.004$ at the location of separation for both the flow over the Boeing speed bump and the flow in the Bachalo-Johnson experiments. For laminar flows, Thwaites also proposed a fit between the shape factor, $H$, and the Holstein-Bohlen parameter, $m$ that allowed computation of the skin friction using the local value of $m$. Using this analysis, the separation boundary for a laminar flow was suggested to be $m=0.09$. Although not shown, we have observed that the two flows considered in this work exhibited separation for different values of $m$. For this reason, the parameter, $m$, can not be used alone to predict flow separation in turbulent flows.  \\

\noindent
Alternatively, it is possible to estimate the value of the Alber separation parameter by using the proposed model. For a locally constant and finite pressure gradient, the proposed model suggests that $\frac{d\theta}{ds}$ is a linear function of $\frac{\theta}{U^2_e} \frac{d P_e}{d s}$. Using the von Karman integral equation, this equation can be further simplified as, 
\begin{equation}
    \frac{C_f}{2} \approx \frac{C_c}{2} \frac{\nu}{U_e \theta} + \frac{C_{Re, \infty}}{2} + \left (\frac{C_m}{2} - \left(2+H\right) \right)  \frac{\theta }{U^2_e} \frac{d P_e}{d s}
    \label{eqn:cfrelation}
    \end{equation}

\noindent
In the limit of an asymptotically large Reynolds number ($Re_{\theta} \rightarrow \infty$) flow nearing separation, an estimate of $H \approx 2.3-2.5$ (see \citet{castillo2004separation} yields that the separation is imminent when
\begin{equation}
\frac{\theta }{U^2_e} \frac{d P_e}{d s} \approx \frac{C_{Re, \infty}}{2 (  (\frac{C_m}{2} - (2+H) )) } \approx   0.003   
\end{equation} which is of the same order of magnitude as the threshold proposed by \citet{alber1971similar}. Equation \ref{eqn:cfrelation} also describes a separation boundary (considered to be the point where the skin-friction crosses zero) for turbulent flows over smooth surfaces for a range of Reynolds numbers as,     
\begin{equation}
    \frac{C_c}{2} \frac{\nu}{U_e \theta} + \left(\frac{C_m}{2} - \left(2+H\right) \right)  \frac{\theta }{U^2_e} \frac{d P_e}{d s} = - \frac{C_{Re, \infty}}{2}
\end{equation}

\noindent
This describes a nonlinear boundary for predicting turbulent separation and reattachment in a space spanned by $1/Re_{\theta}, \; H, \; \mathrm{and} \;  \frac{\delta^{*} }{U^2_e} \frac{d P_e}{d s} $.

 \section{\label{sec:adjointsens} Sensitivity analysis of separation using Alber's parameter}
 
\noindent
The accurate prediction of flows exhibiting separation from mild, adverse pressure gradients remains a pacing item for the development of closure models 
for both Reynolds-averaged Navier-Stokes (RANS) and large-eddy simulations (LES).  While it is known that many of these flows exhibit history effects (potentially originating from the inflow conditions), there are not many computationally efficient ways to quantify the impact of these history effects. Moreover, for a given flow, it is unclear which spatial locations are particularly sensitive with respect to the prediction of the separation point.  The 
following subsections explore the use of the proposed model to quantify these sensitivities.  The following assumes that the point of separation is governed by Alber's parameter, as was motivated in the preceding section.

\subsection{Sensitivity to inflow conditions}
The sensitivity of the Alber parameter (denoted by $J(\theta) = -\frac{\theta}{U_e} \frac{dU_e}{ds}$ here) to the inflow conditions is given by $\frac{dJ}{d\theta_0}$.  

Upon rewriting the ODE model in terms of Alber's parameter, the following expression can be written, 
\begin{equation}
    \theta(s)  = \theta_0 +  \frac{C_c\nu}{2} \int^{s}_0  \frac{1}{U_e \theta} dr - \frac{C_{m}}{2}  \int^{s}_0 J(\theta) dr + \frac{C_{Re, \infty}}{2}s 
\end{equation}
The sensitivity of the momentum thickness due to a perturbation in the inlet momentum thickness ($\theta_0$) can be then measured as 
\begin{equation}
    \frac{d \theta(s)}{d\theta_0} = 1 -  \frac{C_c\nu}{2} \int^{s}_0 \frac{1}{U_e \theta^2} \frac{d\theta}{d\theta_0} dr + \frac{C_m}{2} \int^{s}_0 \frac{1}{U_e}\frac{dU_e}{dr} \frac{d\theta}{d\theta_0} dr 
\end{equation}
where it has been assumed that the inlet is sufficiently far upstream such that the growth of the inlet boundary layer thickness is negligible compared to other terms in the equation. The resulting system can be recast into a separable second-order differential equation relating $\theta \; \mathrm{and} \;  \theta_0$ as
\begin{equation}
    \frac{d^2 \theta(s)}{ds d\theta_0}  +  \left(\frac{C_c\nu}{2U_e \theta^2} - \frac{C_m}{2} \frac{1}{U_e}\frac{dU_e}{ds}\right) \frac{d\theta}{d\theta_0}  = 0
\end{equation}

\noindent
Assuming that Alber's threshold on $J(\theta)$ is approximately accurate (as also shown in Figure \ref{fig:separationptalber}), the sensitivity of $J(\theta)$ at a location of interest $s^*$ (region close to the ``true" point of separation) to inflow conditions is given as 
\begin{equation}
     \frac{dJ}{d\theta_0}  \bigg|_{s^*}  =  \left[\frac{1}{U^2_e}\frac{dP_e}{ds}   e^{-\frac{C_m}{2} \int^{s^*}_0 \frac{1}{U^2_e} \frac{dP_e}{ds} dr } \right] \; \left[e^{-\frac{C_c \nu} {2} \int^{s^*}_0 \frac{dr }{U_e \theta^2}} \right] 
\end{equation}

\noindent
This implies that the value of Alber's parameter is exponentially sensitive (in a Lagrangian sense) to both the pressure gradient and the non-dimensional group $\frac{s^*}{Re_{\theta} \theta }$. The latter carries the group $\frac{s}{\theta} = \frac{U^2_e s}{U^2_e \theta }$ which is a measure of the integrated momentum deficit in the viscous and the momentum transport in an inviscid region. Further, as $s^* \rightarrow \infty $, the asymptotic limit of $\frac{dJ}{d\theta} \rightarrow 0$ or that the parameter is insensitive to the inflow. Secondly, in the limit of infinitely large Reynolds number, $Re_{\theta}$, the parameter (or the location of the separation) becomes purely a function of the pressure gradient as expected. 



\begin{figure}
    \centering
    \includegraphics[width=1.05\textwidth]{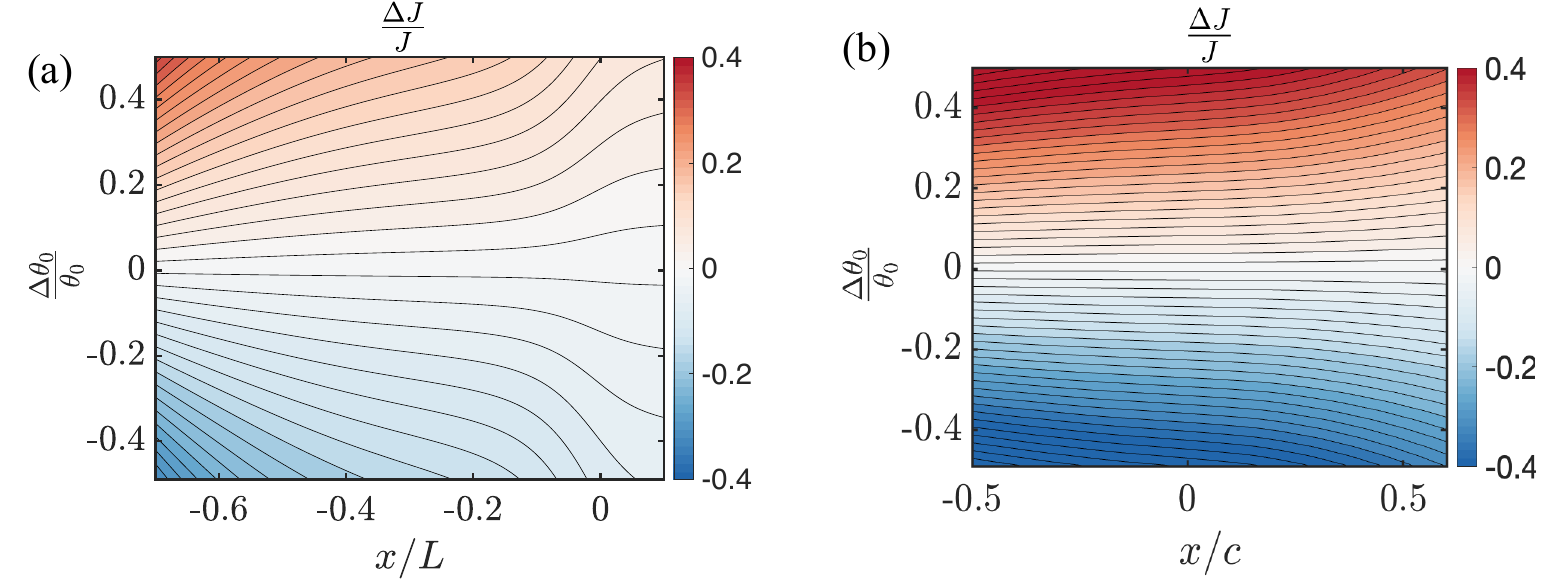}
     \caption{The relative sensitivity of $J = -\frac{\theta}{U_e} \frac{dU_e}{ds} $ to the relative change in inlet momentum thickness ($\frac{\Delta \theta_0}{\theta_0}$) pre-separation for the (a) Boeing speed bump \citep{uzun2021high} and for (b) the transonic Bachalo-Johnson bump \citep{uzun2019wall}. Note that $x/L = 0$ is the position of the apex of the Boeing speed bump, whereas $x/c = 0.5$ is the position of the apex of the transonic bump. }
    \label{fig:adjoint}
\end{figure}

\noindent
Figure \ref{fig:adjoint} shows the sensitivity of $J$ to a perturbation in $\theta_0$ for the SBSE and BJ flows considered herein. Traversing along the x-axis of the plot, the local (in space), relative sensitivity of $J$ to a given perturbation in $\theta_0$ can be quantified ($\theta_0/\theta_0$ represents the number on the ordinate on the left side the figures). Alternatively, Figure \ref{fig:adjoint} can be interpreted as follows: for a given location on the x-axis, the contours are suggestive of the change in $J$ due to a change in $\theta_0$ (plotted on the ordinate on the left side). For the Boeing speed bump, inflow errors (in $\theta$) as large as $ \approx 50\%$ do not appreciably change the value of the Alber's parameter. However, for the Bachalo-Johnson transonic bump, the inflow conditions can affect the momentum thickness far downstream.


 \subsection{  Sensitivity to perturbations in the Holstein-Bohlen parameter }

In scale-resolving simulation paradigms such as wall-modeled large eddy simulations, it is 
difficult to isolate the regions of deficiency of the near-wall model (for predicting the wall shear stress) in non-equilibrium flows, especially in flows undergoing turbulent smooth body separation. The parameter $ \frac{m}{J_{sep}} \frac{dJ_{sep}}{dm}$ is a relative measure of the sensitivity of flow separation to upstream modeling deficiencies (the perturbations in $m$ can be thought of as arising from the near-wall errors in the prediction of wall shear stress). This parameter can be re-expressed as, 
\begin{equation}
    \frac{m}{J_{sep}} \frac{d J_{sep}}{dm} =  \frac{m}{J_{sep}} \frac{dJ_{sep}}{dJ(\theta)} * \frac{dJ(\theta)}{dm}  =\frac{m}{J_{sep}}   \frac{dJ_{sep}}{dJ(\theta)} * \frac{1}{2 Re_{\theta}} = \frac{1}{2} \frac{\theta}{\theta_{sep}} \frac{d \theta_{sep}}{d \theta} = \frac{1}{2} \frac{\theta}{\theta_{sep}} \dfrac{1}
    {\frac{d \theta }{d \theta_{sep}}}
\end{equation}

\begin{figure}
    \centering
    \includegraphics[width=1\textwidth]{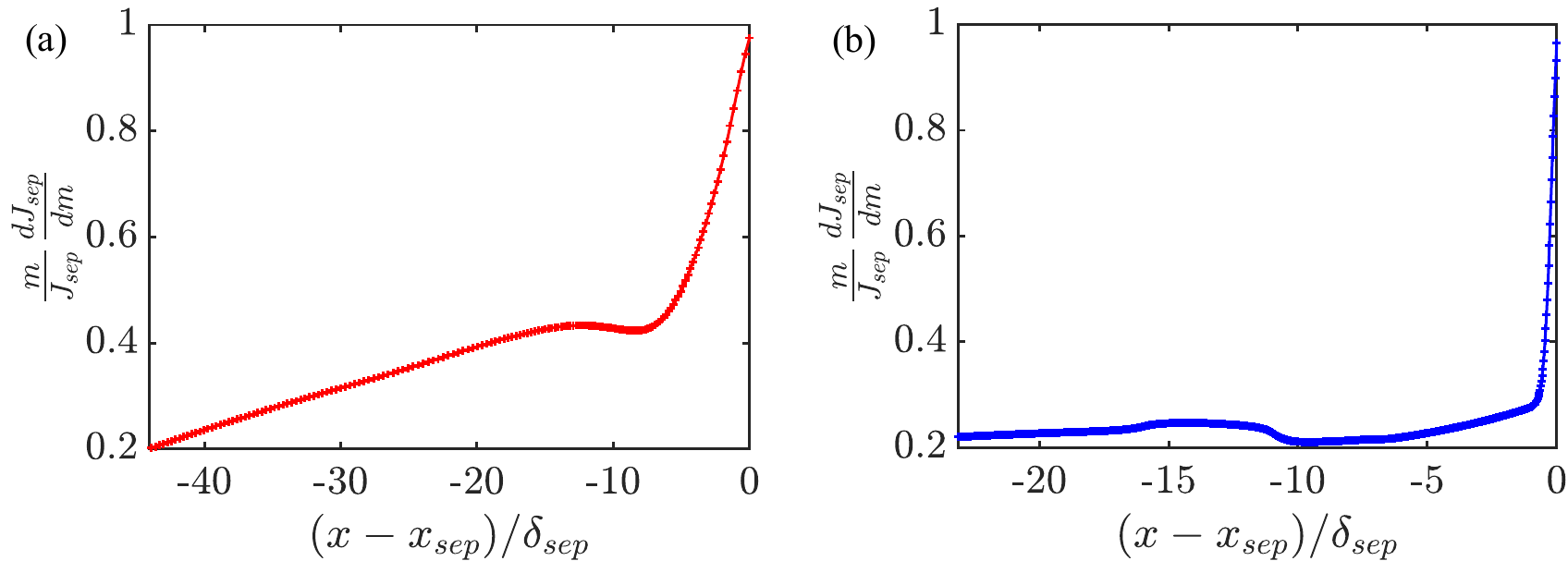}
    
    \caption{The streamwise variation (non-dimensionalized by the boundary layer thickness at the point of separation) of the relative sensitivity of $J_{sep} = -\frac{\theta}{U_e} \frac{dU_e}{ds} $ evaluated at the point of separation to the upstream perturbations in the Holstein-Bohlen parameter, $m$, for (a) Boeing speed bump and (b) transonic Bachalo-Johnson bump. }
    \label{fig:adjoint2}
\end{figure}

\noindent
By perturbing the momentum thickness at the true point of separation, and advancing the proposed model fit for $\theta$ upstream, the sensitivity parameter can be completely determined. Figure \ref{fig:adjoint2} suggests that two different mechanisms are important for two separating flows considered above. For the Boeing speed bump, the sensitivity of Alber's parameter to a perturbation in the Holstein-Bohlen parameter is non-local. Perturbations in the region after the apex (the adverse pressure gradient region) are most important, in that the imminently separating flow responds to those the most. The curve is increasing in the zero pressure gradient and the favorable pressure gradient region, which implies that the closer the perturbations are made to the separation point, the more responsive the tendency of the flow to actually separate in predictive simulations. These results are in agreement with \citet{devenport2022equilibrium} who suggest that the ``history effects" of pressure gradients in a turbulent boundary layer may extend up to 50$\delta$. However, for the transonic bump, the sensitivity is primarily localized in a small region after the apex (in the adverse pressure gradient region), just before the shock. This suggests that the prediction of separation is primarily governed by the shock and the adverse pressure gradient just upstream of it; or that the upstream history or modeling errors are not that significant. These contrasting mechanisms provide clear evidence that flow separation can occur due to different flow development patterns, and future modeling efforts may need to account for these effects.

 \section{\label{sec:Conclusion} Concluding remarks}
This work presents an extension of the method of Thwaites for determining the momentum thickness for a turbulent boundary layer under the action of pressure gradients. In the limit of large Reynolds numbers, a linear model is hypothesized, such that the Reynolds number and pressure gradient dependence on the growth rate of the momentum thickness can be superposed. A fit for the model coefficients is found using recent high-fidelity simulation data for various boundary layers ranging from Reynolds number, $Re_{\tau} \approx [100, 4000]$ and Clauser parameter, $\beta, \approx [-6, 10]$. The model fit predicts the growth of the momentum thickness, even up to the point of separation. The fits utilized in the model can then be used to derive a condition for imminent separation at high Reynolds numbers, with similar threshold values as compared to previously proposed values based on empirical data. An application of this extension of Thwaites method is demonstrated by estimating the importance of history effects on separation. Specifically, the importance of upstream history effects for the Boeing speed-bump flow is quantified. For a transonic bump flow, it is shown that only local perturbations just upstream of the shock location cause significant changes to the separation location. 

\appendix

 \section{\label{sec:deltastareqn} Relation between of $\delta$, $\delta^{*}, \; \mathrm{and} \; \theta$ }
  For laminar flows, the solution to Thwaites method for $\theta$ and the universal correlation between $m$ and the shape factor $H$ provides a $\delta^{*}$ which is useful for iteratively updating the ``inviscid geometry'' that is used for computing the freestream profiles. The proposed model in this work provides a good fit for $\theta$, but a fit for $\delta^{*}$ is needed as well for iterative deployment with a potential flow solver. The analytical expression for the displacement thickness can be derived from the continuity equation as, 
  \begin{equation}
      \frac{d \delta^{*}}{d s} = \frac{V_e}{U_e} + \frac{1}{U^2_e} \frac{d U_e}{ds} \int^{\delta}_0 U dn 
  \label{eqn:integral}
  \end{equation}

\noindent
where $\delta$ is the thickness of the boundary layer. For general flows, the integral in Equation \ref{eqn:integral} is the measure of the mass flow rate inside the boundary layer, and is dependent on the local flow conditions (such as $Re_{\tau}$ and pressure gradient $\frac{dP_e}{ds}$) and is unknown ``a priori". For given values of $\delta$ and the flow variables at the edge of the boundary layer $U_e$, $V_e$, and $P_e$, the growth rate of the displacement thickness is given as
 \begin{equation}
         \frac{d \delta^{*}}{d s} = \frac{V_e}{U_e} - \frac{1}{U^2_e}\frac{d U_e}{ds} \int^{s}_{s_0} \frac{V_e(r)}{U_e(r)}U_e (r) dr, 
    \label{eqn:ddeltastardx}
 \end{equation}
 where $s_0$ is the streamwise location at which the flow can be first considered fully turbulent, within the boundary layer, and $s$ is the location of interest. From geometrical arguments, the ratio $V_e/U_e$, can be approximately related to the growth of the boundary layer thickness as
  \begin{equation}
     \frac{V_e}{U_e} \approx \frac{d \delta}{ds}
 \end{equation}
 Thus, the growth of the boundary layer affects the growth rate of the displacement thickness linearly as follows, 
  \begin{equation}
         \frac{d \delta^{*}}{d s} = \frac{d \delta}{ds} - \frac{1}{U^2_e}\frac{d U_e}{ds} \int^{s}_{s_0} \frac{d \delta}{dr} U_e (r) dr 
 \end{equation}
 
\begin{figure}
    \centering
    {\includegraphics[width=0.5\textwidth]{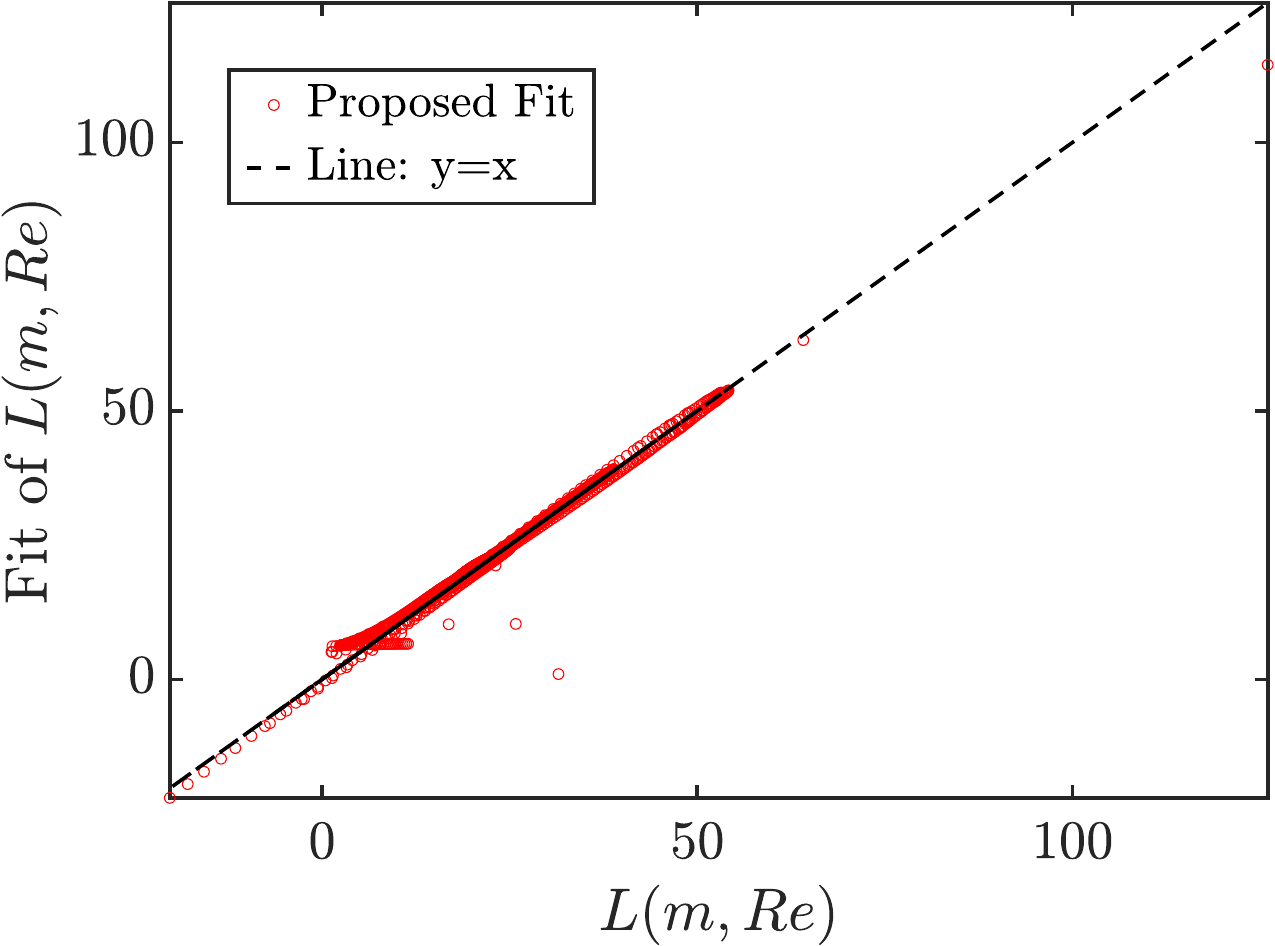}}

    \caption{ The quality of the fit between the exact value of $L(m, Re)$ and that obtained from the proposed Equation\ref{eqn:ddeltadx}.  The red symbols contain data from the zero-pressure gradient boundary layer of \citet{eitel2014simulation}, the five adverse pressure gradient boundary layers of \citet{bobke2017history}, and the three boundary layers from the NACA airfoils \citep{vinuesa2017revisiting,tanarro2020effect}.  }
    \label{fig:ddeltadx}
\end{figure}

Finally, a linear relationship between $V_e/U_e$, $L$ and $m$ is fitted from the simulation data as
\begin{equation}
    L(m, Re) =  \frac{U_e}{\nu } \frac{d \theta^2}{d s} \approx 5 + 8 m - 200 \frac{d \delta }{d s} + 200 \frac{d \delta{zpg,corr}} {ds}
\end{equation}
which implies that,  \begin{equation}
   \frac{d \delta }{d s} \approx \frac{5}{200} + \frac{8}{200} m - \frac{U_e}{200 \nu } \frac{d \theta^2}{d s}   +  \frac{d \delta{zpg,corr}} {ds}
\label{eqn:ddeltadx}
\end{equation}
where $\delta_{zpg,corr}/s = 0.16 Re^{-1/7}_s = 0.16 (U_e (s) s / \nu )^{-1/7} $. 
The quality of this fit is verified in Figure \ref{fig:ddeltadx} by comparing the exact values of $L(m, Re)$ obtained from the datasets considered in this work, and from Equation \ref{eqn:ddeltadx}. With the two proposed fits in this work, the displacement thickness can be determined along the streamwise coordinate using Equations \ref{eqn:ddeltastardx} and \ref{eqn:ddeltadx}. 

\section*{Acknowledgments} 
This work was supported by NASA’s Transformational Tools and Technologies project under Grant No. \#80NSSC20M0201 and by Boeing Research \& Technology. K.P.G. is supported by the Exascale Computing Project (Grant17-SC-20SC). 

This work was authored in part by the National Renewable Energy Laboratory, operated by Alliance for Sustainable Energy, LLC, for the U.S. Department of Energy (DOE) under Contract No. DEAC3608GO28308. The views expressed in the article do not necessarily represent the views of the DOE or the U.S. Government. The U.S. Government retains and the publisher, by accepting the article for publication, acknowledges that the U.S. Government retains a nonexclusive, paid-up, irrevocable, worldwide license to publish or reproduce the published form of this work or allow others to do so, for the U.S. Government purposes. 

\vspace{10pt}
\noindent
\textbf{Declaration of Interests}: The authors report no conflict of interest.

\bibliographystyle{jfm}

\end{document}